\documentclass[aps,prl,twocolumn,groupedaddress]{revtex4-1}

\usepackage{graphicx} 
\usepackage{color}

\begin{document}


\title{Multiple Magnetization Plateaus and the Magnetic Structures in $S=1/2$ Heisenberg Model on the Checkerboard Lattice}


\author{Katsuhiro Morita}
\email[e-mail:]{morita@cmpt.phys.tohoku.ac.jp}
\author{Naokazu Shibata}
\affiliation{Department of Physics, Tohoku University, Aoba-ku, Sendai 980-8578 Japan}


\date{\today}

\begin{abstract}
We study the ground state of $S = 1/2$ Heisenberg model on the checkerboard lattice in a magnetic field by the density matrix renormalization group (DMRG) method with the sine-square deformation. We obtain magnetization plateaus at $M/M_{\rm sat}=$0, 1/4, 3/8, 1/2, and 3/4 where $M_{\rm sat}$ is the saturated magnetization. The obtained 3/4 plateau state is consistent with the exact result, and the 1/2 plateau is found to have a four-spin resonating loop structure similar to the six-spin loop structure of the 1/3 plateau of the kagome lattice. Different four-spin loop structures are obtained in the 1/4 and 3/8 plateaus but no corresponding states exist in the kagome lattice. 
The 3/8 plateau has a unique magnetic structure of three types of four-spin local quantum states in a $4\sqrt{2}\times2\sqrt{2}$ magnetic unit cell with a 16-fold degeneracy.
\end{abstract}


\maketitle
Frustrated quantum spin systems exhibit unusual
quantum phenomena such as the formations of valence bond solid (VBS), spin-nematic, and quantum spin liquid (QSL) as a result of competing quantum fluctuations \cite{Mila,Balents}. 
Application of an external magnetic field complicates the situation further with the increase in uniaxial anisotropy, and it sometimes leads to a phase transition to a stable quantum state with a jump or
cusp in the magnetization curve at zero temperature. 
Recently observed 1/3 magnetization plateau in the triangular lattice of $S=1/2$ quantum spins \cite{tri1,tri2,tri3} is a typical example, in which up-up-down spin structure emerges with a finite excitation gap \cite{tri4,tri5,tri6,tri-sq}. 
Similar 1/2 plateau is obtained in the $J_1$-$J_2$ square lattice \cite{tri-sq,sq,sq2} and novel multiple magnetization plateaus are realized in the Shastry-Sutherland lattice \cite{ss1,ss2,ss3,ss4}. 
In the kagome lattice, 0, 1/9, 1/3, 5/9, and 7/9 plateaus are predicted by the DMRG method
with the sine-square deformation \cite{kago1}.

\begin{figure}
\includegraphics[width=27mm]{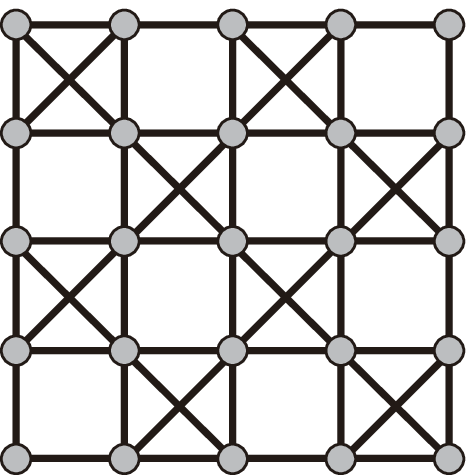}
\caption{The checkerboard lattice.
\label{checkerlattice}}
\end{figure}

\begin{figure}
\includegraphics[width=64mm]{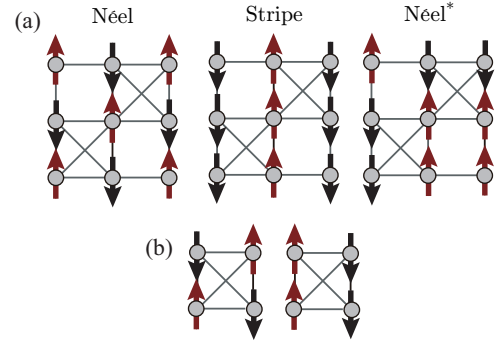}
\caption{(a) The ground spin configurations of the classical Heisenberg model on the checkerboard lattice at zero magnetic field. (b) Possible ground state local spin configurations. The total moment of four spins connected by diagonal interactions is zero.
\label{classical}}
\end{figure}

The checkerboard lattice shown in Fig.~\ref{checkerlattice} is another example of
frustrated system and referred to as two-dimensional pyrochlore lattice. Its classical ground state at zero magnetic field is obtained when the magnetic moments of four spins connected by diagonal interactions are canceled as shown in Fig.~\ref{classical}. 
This means the presence of a macroscopic degeneracy in the ground state
as in the cases of the kagome lattice and the $J_1$-$J_2$ square lattice at 
$J_2/J_1$= 0.5. 
In contrast, $S=1/2$ quantum spins on the checkerboard lattice have stable quantum ground state as in the kagome lattice \cite{kago1,kago2,kago3} and Shastry-Sutherland lattice \cite{ss1,ss2,ss3,ss4}. 
Indeed previous studies on the checkerboard lattice indicate that the ground state is a plaquette valence-bond crystal (PVBC)
with a large spin singlet-triplet gap,  $\Delta_{\rm st} \approx 0.6J$ \cite{ch1,ch2} and the  3/4 plateau is realized just before the saturation 
magnetic field \cite{chex} as in the case of the 7/9 plateau in the kagome lattice \cite{kagoex}.

Although the ideal checkerboard lattice with only one exchange energy $J$ has not been synthesized, similar model substances are present \cite{chexp1,chexp2}.
The checkerboard lattice is obtained by replacing the position of anions and cations in the CuO$_2$-plane structure. 
When the nearest-neighbor exchange interaction $J_1$ and the two next-nearest neighbor exchange interactions $J_2$ and $J_2^\prime$ satisfy the condition that $J_2 \approx J_1$ and $J_2^\prime \approx$ 0,  the Heisenberg model with only one exchange energy $J$ on the checkerboard lattice is realized.

In this article, we examine the ground state of the $S$=1/2 Heisenberg model on the checkerboard lattice by large-scale DMRG calculations and clarify many stable quantum states realized in a magnetic field.
The Hamiltonian of this model is defined as
\begin{eqnarray} 
H &=& J\sum_{\langle i,j \rangle } \mathbf{S}_i \cdot \mathbf{S}_j - h\sum_i S^{z}_i,
\end{eqnarray}
where the sum of the first term is for pairs of spins on all bonds written in Fig~\ref{checkerlattice}. 

\begin{figure}
\includegraphics[width=82mm]{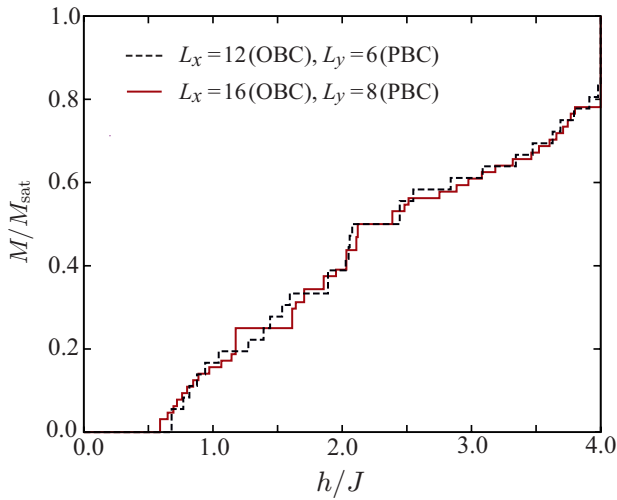}
\caption{$M/M_{\rm sat}$ vs $h/J$ in cylindrical systems. 
The red solid and black dashed lines correspond to the system of $L_x$ = 16, $L_y$ = 8 and $L_x$ = 12, $L_y$ =  6, respectively.
\label{simpleM-ch}}
\end{figure}

\begin{figure}
\includegraphics[width=70mm]{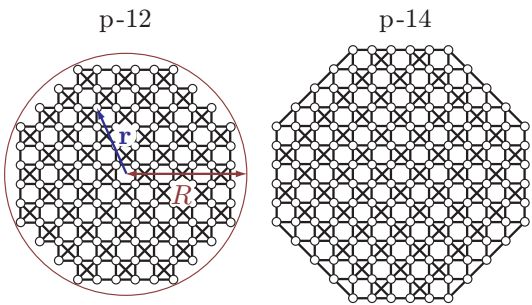}
\caption{Two octagonal clusters of the checkerboard lattice used in the present grand canonical SSD method.
\label{ssdc}}
\end{figure} 
Before we analyze the detailed properties of this model we first briefly examine the magnetization process in the two finite cylinders of length $L_x$ = 12 and 16 with the open boundary conditions (OBC)  and $L_y$= 6 and 8 with the periodic boundary conditions (PBC), respectively.
The magnetization curves obtained by the DMRG method are shown in Fig.~\ref{simpleM-ch}, where we find severe size dependence of the magnetization plateaus.
This size dependence is mainly caused by the periodic boundary conditions on $L_y$ and systematic calculation for large $L_y$ beyond the correlation length of the ground state is indispensable to confirm the presence of the plateaus.
We also need to remove the effect of open boundary conditions on $L_x$, which makes edge spins contribute to an artifactual shift of the plateaus.
For these reasons, the analysis on the bulk properties from the finite systems of available sizes is difficult under usual boundary conditions.

To overcome this difficulty, here we use the ground canonical analysis with the sine-square deformation (SSD) that has been developed recently \cite{ssd1,ssd2,ssd3,ssd4}.
The SSD deforms the original Hamiltonian of Eq. (1) to that locally rescaled by the function $f(\mathbf{r})$ as
\begin{eqnarray} 
H &=& J\sum f\left(\frac{\mathbf{r}_i+\mathbf{r}_j}{2}\right) \mathbf{S}_i \cdot \mathbf{S}_j - h\sum_i f(\mathbf{r}_i)S^{z}_i,
\end{eqnarray}
where $f(\mathbf{r})$ is a decreasing function of $|\mathbf{r}|$ from the center of the system defined by $f\left(\mathbf{r}\right) = \frac{1}{2}\left[1+\textrm{cos}\left(\frac{\pi |\mathbf{r}|}{R}\right)\right]$ and it vanishes at the distance $R$ from the center, which corresponds to the boundary of the system as shown in Fig.~\ref{ssdc}.
This technique is known to reduce the finite size effects and reasonably reproduce the correct bulk properties \cite{ssd4}.

\begin{figure}
\includegraphics[width=82mm]{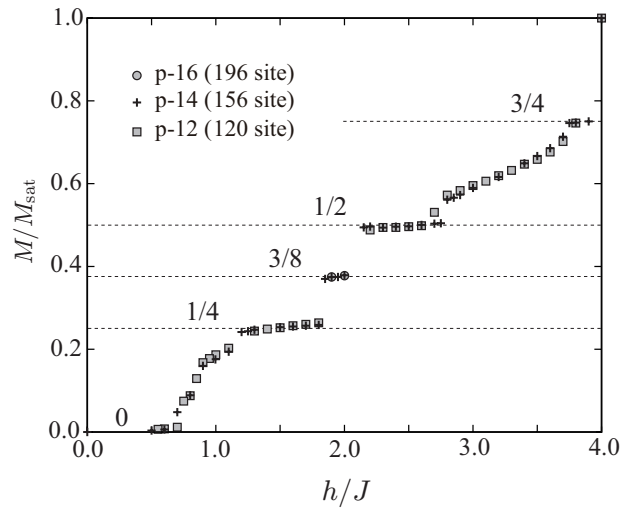}
\caption{$M/M_{\rm sat}$ vs $h/J$ obtained by the grand canonical SSD method. The numbers on the plateaus correspond to $M/M_{\rm sat}$.
\label{ssdM}}
\end{figure} 
\begin{figure}
\includegraphics[width=82mm]{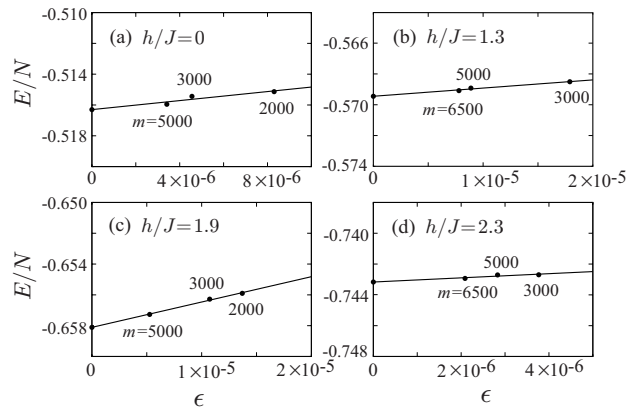}
\caption{Ground-state energy per site, $E/N$, as a function of the truncation error, $\epsilon$, obtained by the DMRG
method with the SSD in the p-14 cluster at $M/M_{\rm sat}$= (a) 0, (b) 1/4, (c) 3/8, and (d) 1/2. The numbers of keeping states, $m$, in the DMRG are shown in the figures.
\label{eps}}
\end{figure} 
In the present study we use the three octagon clusters p-12, p-14 and p-16 as shown in Fig.~\ref{ssdc}, where 12, 14 and 16 are the largest number of spins aligned in one direction. 
In our DMRG calculations, the number of states, $m$, retained in each block is 1000 -- 6500.
The truncation error is less than $2.0\times 10^{-5}$ at low magnetic fields and less than $5.0\times 10^{-6}$ in other cases. 
The accuracy of the present calculation is confirmed at $h/J$= 3.9, where the exact energy of the 3/4 plateau state is reproduced within 0.35\% with the number of keeping states $m=1000$.
The $m$ dependence of the ground state energy shown in Fig.~\ref{eps} 
confirms that the energy difference from the extrapolated value in the limit of $\epsilon \rightarrow 0$ is well scaled by the truncation error and this difference is typically less than 0.1\% for $m$ $\ge$ 5000 states. 
\begin{figure}
\includegraphics[width=78mm]{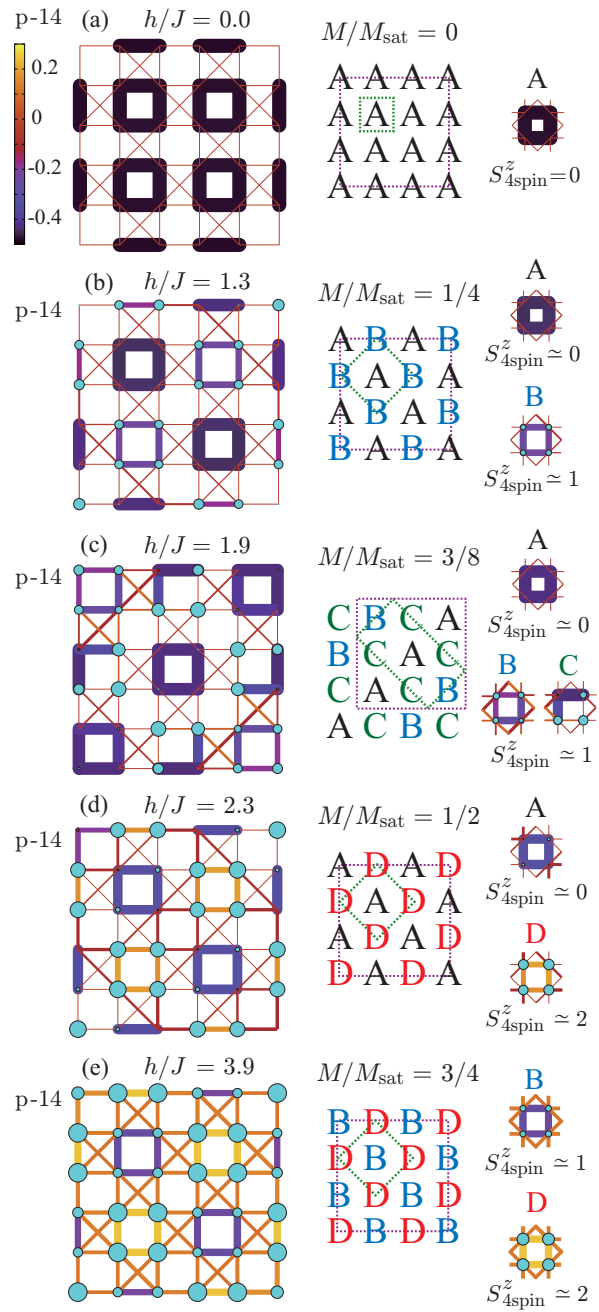}
\caption{The left side figures show the nearest- and next-nearest-neighbor correlations
$\langle \mathbf{S}_i\cdot\mathbf{S}_j\rangle$ and the local magnetizations $\langle S^z_i\rangle$ of 6 $\times$ 6 sites around the center of p-14 cluster in the plateau phase at $M/M_{\rm sat}$= (a) 0, (b) 1/4, (c) 3/8, (d) 1/2, and (e) 3/4. The thickness and color of the lines represent the magnitude and sign of the correlations. The diameter and the color of the circles on the lattice represent the magnitude of the $\langle S^z_i\rangle$ and the sign (blue and red are positive and negative), respectively. The figures in the middle show the patterns of magnetic structure. The letters, A, B, C, and D correspond to a four-spin quantum state of $S^z_{\rm 4spin} \simeq$ 0, 1, 1, and 2, respectively, where $S^z_{\rm 4spin}$ is the total $S^z$ of the four-spin state. B and C have different spin structures. 
The purple and green dashed lines represent 6 $\times$ 6 sites of the left side
figures and the magnetic unit cell, respectively.
\label{ss1}}
\end{figure} 

We first show the magnetization $M/M_{\rm sat}$ in Fig.~\ref{ssdM}, which is evaluated within a magnetic unit cell in the central 6 $\times$ 6 spins in the octagonal clusters. We clearly find five magnetization plateaus at $M/M_{\rm sat}$ = 0, 1/4, 3/8, 1/2, and 3/4.
In the following we investigate the detailed real space structure of local magnetizations $\langle S^z_i\rangle$ and the nearest- and next-nearest neighbor correlation functions $\langle \mathbf{S}_i\cdot\mathbf{S}_j\rangle$.

Figure~\ref{ss1} shows $\langle S^z_i\rangle$ and $\langle \mathbf{S}_i\cdot\mathbf{S}_j\rangle$ in the plateau states. We find that each plateau has a characteristic four-spin resonating loop structure.
Since the exchange energy of these states is determined by the local spin correlations $\langle \mathbf{S}_i\cdot\mathbf{S}_j\rangle$, the smallest resonating loop is realized to enhance the short range quantum fluctuations on the position of four spins which are not connected by competing diagonal exchange interactions.  
The ground state at zero magnetic field is well characterized by PVBC consisting of four-spin loops as shown in Fig.~\ref{ss1}(a). The spin singlet-triplet gap, $\Delta_{\rm st}$, is estimated to be 0.6$J$ from the magnetization curve in Fig.~\ref{ssdM}, which is consistent with the previous calculations \cite{ch1,ch2}.

\begin{figure}
\includegraphics[width=70mm]{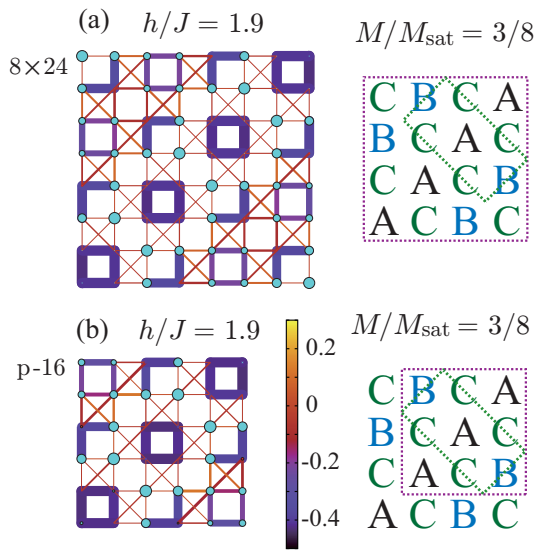}
\caption{The nearest- and next-nearest-neighbor correlations $\langle \mathbf{S}_i\cdot\mathbf{S}_j\rangle$ and the local magnetizations $\langle S^z_i\rangle$ of (a) 8 $\times$ 8 sites around the center of 8 (PBC) $\times$ 24 (SSD) cylinder and (b) 6 $\times$ 6 sites around the center of p-16 cluster in the 3/8 plateau. The right side figures show the patterns of magnetic structure in the 3/8 plateau. 
The purple dashed lines represent 8 $\times$ 8 unit of (a) and 6 $\times$ 6 unit of (b),  respectively, and the green dashed line represent magnetic unit cell.
\label{1ssd}}
\end{figure} 

In the plateaus at $M/M_{\rm sat}$ = 1/4, 1/2, and 3/4, we find two types of  four-spin loops in the magnetic unit cell of eight spins.
In the 1/2 plateau, the magnetic structure has $2\sqrt{2}\times2\sqrt{2}$ unit structure with two types of four-spin loops of $S^z_{\rm 4spin}$ $\simeq$ 0 and 2 as shown in Fig~\ref{ss1}(d) where $S^z_{\rm 4spin}$ represents total $S^z$ of four spins in the loop. This magnetic structure is different from the uuud structure realized in the $J_1$-$J_2$ square lattice \cite{tri-sq,sq,sq2}, and this difference in the ground state between the checkerboard and the $J_1$-$J_2$ square lattice is similar to the difference between those of the kagome and the triangular lattice in the 1/3 magnetization plateau, where hexagonal singlet loops are realized in the kagome lattice while uud state appears in the triangular lattice. 

Similar spin structure to that of the 1/2 plateau is realized in the 3/4 plateau, where four-spin loops composed of $S^z_{\rm 4spin}$ $\simeq$ 1 and 2 appear as shown in Fig~\ref{ss1}(e).  Thus the 1/2 and 3/4 plateaus correspond to the 1/3 and 7/9 plateaus in the kagome lattice \cite{kago1} and the 1/3 and 2/3 plateaus in the square-kagome lattice \cite{sq-kago1,sq-kago2}, respectively.
The common feature of these ground state is the presence of closed resonating loops surrounded by almost fully polarized spins each of them is nearly independent. The ground state is then composed of local quantum singlet, triplet, or quintet states of closed loops of four- or six-spins. Since the exchange energy of even number quantum spins on a closed loop is lower for smaller loop, resonating loops of the smallest even number of spins is stabilized in quantum systems under the competition with the Zeeman energy, which makes polarized domains.

In the 1/4 plateau, the spin structure is characterized by the resonating loops of $S^z_{\rm 4spin}$ $\simeq$ 0 and 1 as shown in Fig~\ref{ss1}(b). Although each four-spin loop is not surrounded by fully polarized spins, it is almost independent of the others.
This kind of magnetic structure is not found in the kagome lattice, and square-kagome lattice 
but found in the $1/5$-depleted Heisenberg square lattice with frustration whose Hamiltonian has a unit of local square structure without diagonal interactions \cite{15sq} as in the checkerboard lattice. 

The 3/8 plateau state is characterized by an unique magnetic structure as shown in Fig.~\ref{ss1}(c).
The magnetic unit cell includes three types of four-spin units one of which has $S^z_{\rm 4spin}$ $\simeq$ 0 and the other two have $S^z_{\rm 4spin}$ $\simeq$ 1.
Although only 6 $\times$ 6 sites around the center of the p-14 octagonal cluster are presented in Fig.~\ref{ss1}(c), this result suggests a periodic structure of 8 $\times$ 8 spins as shown in the right side panel of Fig.~\ref{1ssd}(a). 
Because of the large magnetic unit cell of the ground state, we do not obtain the 3/8 plateau in the p-12 cluster.
We therefore calculate the ground state correlations $\langle \mathbf{S}_i\cdot\mathbf{S}_j\rangle$ and local magnetizations $\langle S^z_i\rangle$ in the p-16 cluster and 8 (PBC) $\times$ 24 (SSD) cylinder at $h/J=1.9$.
The obtained results are consistent with that of the p-14 cluster as shown in Fig~\ref{1ssd}. 
Since three types of four-spin units written by A, B, and C in the right side of Fig.~\ref{1ssd} are only weakly correlated and almost independent with each other, this magnetic structure is expected to be stable even in the thermodynamic limit. 
Thus we conclude that the 4$\sqrt2$$\times$2$\sqrt2$ unit cell with three types of four-spin units is the intrinsic magnetic structure of the 3/8 plateau.
We note that the ground state has 16-fold degeneracy coming from the inversion symmetry and the translation symmetries along $x$, $y$ and the diagonal directions.

We next discuss the stability of the 1/3 plateau shown in Fig.~\ref{simpleM-ch} which is obtained in a finite small system of length $L_x = 12$ and $L_y = 6$. Although the 1/3 plateau is quite clear, we find strong anisotropic correlations in the plateau state peculiar to the system of $L_y=6$ with PBC.
In fact, similar result is obtained in twice large system of $L_x$ = 24 (SSD) and $L_y$ = 6 (PBC) where the nearest-neighbor correlations $\langle \mathbf{S}_i\cdot\mathbf{S}_j\rangle$ are -0.06 and -0.26 in the direction of $x$ and $y$, respectively.
These anisotropic correlations are induced by the six-spin loop structure of $S^z_{\rm 6spin} \simeq$ 1 along $y$ direction that is characteristic to the systems of $L_y$ = 6. Since the 1/3 plateau is not stabilized in the system with open boundaries and in octagonaler clusters, we conclude that the 1/3 plateau is absent in the bulk system.

We find a kink like anomaly in Fig.~\ref{ssdM} at $h/J$ $\approx$ 0.9. 
This may be a sign of 1/8 plateau though we need further investigation.
The magnetic structures of all plateaus show that the same type of four-spin unit is diagonally aligned. 
In the 3/4 plateau however various placement are possible depending on the size and boundary conditions of the system \cite{3/4}.
This is explained by localized one-magnon states, where each magnon is localized in a loop of four-spins surrounded by fully polarized spins.

We finally compare the magnetic structures of the checkerboard lattice with those of the $J_1$-$J_2$ square lattice. Although the ground state of the $J_1$-$J_2$ square lattice at $J_2/J_1 \approx 0.55$ is expected to be PBVC \cite{JJ}, magnetization plateaus are not obtained except for 1/2 plateau, where uuud structure is observed.
The origin of this difference is coming from the presence of the ferromagnetic diagonal spin-spin correlations of the four-spin loops of $S^z_{\rm 4spin}$ $\simeq$ 0 and 2. 
In the checkerboard lattice, these four-spin resonating loops are stable, since missing diagonal interaction within the loop keeps the energy of each four-spin state,
while in the $J_1$-$J_2$ square lattice, these loops are unstable due to the presence of the diagonal antiferromagnetic interactions, $J_2$.
In the uuud structure, however, 
the number of interacting up-up spin pair in each square unit is the same to the number of up-down spin pair both for the  $J_1$-$J_2$ square lattice and the checkerboard lattice, and there is no penalty for $J_2$ in the $J_1$-$J_2$ square lattice. 
In quantum states with short range antiferromagnetic correlations, the diagonal antiferromagnetic interactions $J_2$ in the $J_1$-$J_2$ square lattice compete with the short range correlations of the nearest-neighbor spin pairs connected by $J_1$ and thus the nearly classical uuud structure is relatively stable in the $J_1$-$J_2$ square lattice \cite{tri-sq,sq,sq2}.

Finally we comment on the presence of the similar plateaus in some related models. The similarity between the checkerboard lattice and the 1/5-depleted square lattice suggests the presence of the plateau in the 1/5-depleted square lattice that corresponds to the 3/8 plateau with three types of local four-spin units. We expect that such frustrated lattices composed of square units without diagonal interactions have similar multiple magnetization plateaus characterized by the four-spin resonating loop structures.

\section{acknowledgment}
The present work was supported by a Grant-in-Aid for Scientific Research (No. 26400344) from JSPS.

\end{document}